\documentstyle[preprint,tighten,aps,floats,psfig,epsfig]{revtex}

\def\ub{{\overline{u}}}
\def\vb{{\overline{v}}}

\begin{document}

\draft
\title{
Critical thermodynamics of three-dimensional chiral model for 
$N > 3$}

\author{
P. Calabrese${}^{1}$, 
P. Parruccini${}^2$, and 
A. I. Sokolov${}^3$, 
}

\address{$^1$Scuola Normale Superiore and INFN,
Piazza dei Cavalieri 7, I-56126 Pisa, Italy.}

\address{$^{2}$Dipartimento di Fisica dell' Universit\`a di 
Pisa and INFN, Via Buonarroti 2, I-56100 Pisa, Italy. }

\address{$^3$Department of Physical Electronics, 
Saint Petersburg Electrotechnical University, 
Professor Popov Street 5, St. Petersburg 197376, Russia. \\
{\bf e-mail: \rm 
{\tt calabres@df.unipi.it},
{\tt parrucci@df.unipi.it},
{\tt ais@sokol.usr.etu.spb.ru},
}}

\date{\today}

\maketitle

\begin{abstract}
The critical behavior of the three-dimensional $N$-vector chiral 
model is studied for arbitrary $N$. The known six-loop 
renormalization-group (RG) expansions are resummed using the Borel 
transformation combined with the conformal mapping and Pad\'e 
approximant techniques. Analyzing the fixed point location and the 
structure of RG flows, it is found that two marginal values of $N$ 
exist which separate domains of continuous chiral phase transitions 
$N > N_{c1}$ and $N < N_{c2}$ from the region $N_{c1} > N > N_{c2}$ 
where such transitions are first-order. Our calculations yield 
$N_{c1} = 6.4(4)$ and $N_{c2} = 5.7(3)$. For $N > N_{c1}$ the structure 
of RG flows is identical to that given by the $\epsilon$ and $1/N$ 
expansions with the chiral fixed point being a stable node. For 
$N < N_{c2}$ the chiral fixed point turns out to be a focus having   
no generic relation to the stable fixed point seen at small 
$\epsilon$ and large $N$. In this domain, containing the physical 
values $N = 2$ and $N = 3$, phase trajectories approach the fixed 
point in a spiral-like manner giving rise to unusual crossover 
regimes which may imitate varying (scattered) critical exponents 
seen in numerous physical and computer experiments.  
\end{abstract}


\section{Introduction}

The critical behavior of three-dimensional frustrated spin models 
with noncollinear order has been the subject for intensive experimental 
and theoretical investigations, since they are expected to belong to 
a new universality class with unusual values of critical exponents 
\cite{Kawamura-98,cp-97,rev-01}. Physically, noncollinear or canted 
order is a consequence of frustration due to either the special 
geometry of the lattice or the competition of different kinds of 
interactions. Examples of the first kind are the three-dimensional 
stacked triangular antiferromagnets (STA) in which magnetic ions are 
located at each site of a stacked triangular lattice~
(e.g. CsMnBr$_3$, VCl$_2$, VBr$_2$). The second case is represented 
by helimagnets like the rare-earth Ho, Dy, Tb, in which magnetic 
spirals are formed along certain directions of the lattice. 

Although Monte Carlo simulations \cite{MC} and experimental 
studies \cite{exp} apparently give contradicting results~(supporting 
either a continuous phase transition with new critical exponents or 
a weak first order one), a quite definite description of the 
critical behavior has been reached in the fixed dimension $D = 3$ 
field-theoretical (FT) approach~\cite{prv-01p,cps-02}. In more detail, 
these and other relevant issues are reviewed in Ref.~\cite{rev-01}, 
where one can find a complete list of references; we mention here 
only the very recent studies.

Field-theoretical investigations of such systems are based on the 
O$(N) \times $O$(M)$ symmetric Landau-Ginzburg-Wilson Hamiltonian~
\cite{Kawamura-98,Kawamura-88}
\begin{equation}
{\cal H} = \int d^D x 
 \left\{ {1\over2}
      \sum_{a} \left[ (\partial_\mu \phi_{a})^2 + r \phi_{a}^2 
\right] 
+ {1\over 4!}u_0 \left( \sum_a \phi_a^2\right)^2 
+ {1\over 4!}  v_0 
\sum_{a,b} \left[ ( \phi_a \cdot \phi_b)^2 - \phi_a^2\phi_b^2\right]
             \right\},\label{LGWH}
\end{equation}
where $\phi_a$ ($1\leq a\leq M$) are $M$ sets of $N$-component 
vectors. Frustrated systems with noncollinear ordering are described 
by the Hamiltonian (\ref{LGWH}) with $M=2$ and $v_0>0$. Negative 
values of $v_0$ correspond to simple ferromagnetic or 
antiferromagnetic ordering and to magnets with sinusoidal spin 
structures \cite{Kawamura-88}. The case $M=3$ corresponds to 
frustrated antiferromagnets with nonplanar spin ordering 
\cite{Kawamura-86}. Moreover, recently the Hamiltonian (\ref{LGWH}) 
has received a lot of attentions because it describes also the 
quantum critical behavior in some Mott insulators \cite{sachdev-02}. 

Fixing $M=2$, the physically relevant cases of frustrated XY and 
Heisenberg models are described by the Hamiltonian (\ref{LGWH})
with $N=2$ and $N=3$ respectively. Field-theoretical studies in the 
framework of $\epsilon = 4 - D$ \cite{Kawamura-98,asv-95,prv-01n}, 
$1/N$ \cite{Kawamura-98,prv-01n,g-02n,g-02p}, and 
$\tilde{\epsilon}=D-2$\cite{adj-90} expansions as well as in fixed 
dimensions $D=2$ \cite{cp-01,cops-02} and $D=3$ 
\cite{as-94,lsd-00,prv-01p,cps-02} reveal that the existence and 
the stability properties of the fixed points depend on $N$ and on 
the spatial dimensionality.
Within the $\epsilon$ expansion, for sufficiently large values of 
$N$ the renormalization group~(RG) equations possess four fixed points: 
the Gaussian fixed point ($v_0=u_0=0$), the O$(2N)$-symmetric ($v_0=0$) 
one and two anisotropic fixed points located in the region $u_0,v_0>0$ 
and usually named chiral and antichiral. The chiral fixed point is 
the only stable one. There is a critical dimensionality $N_c$ where 
the chiral and antichiral fixed points coalesce and disappear for 
$N<N_c$. In the last case, under the absence of any stable fixed point 
the system is expected to undergo a weak first-order phase transition, 
since the associated RG flows run away from the region of stability 
of the fourth-order form in the free-energy expansion. The three-loop 
estimate of $N_c$ obtained in the framework of the $\epsilon$ 
expansion technique is \cite{asv-95}
\begin{equation} 
N_c= 21.8-23.4 \epsilon +7.1 \epsilon^2+ O(\epsilon^3)\, ,
\label{NCe}
\end{equation}
that, after an appropriate resummation, results in $N_c > 3$ in 
three dimensions. This inequality leads to the conclusion that 
for the physical models with $N=2, 3$ the three-dimensional chiral 
transition is first-order, as corroborated also by some other RG 
studies~\cite{as-94,tdm-00}. To the contrary, both for $N=2$ and $N=3$ 
the highest-order six-loop calculations in three dimensions 
\cite{prv-01p} reveal a strong evidence for a stable chiral fixed 
point that, however, is not related to its counterpart found earlier 
within the $\epsilon$ expansion. Furthermore, in a recent work 
\cite{cps-02} we claimed that both in two and three dimensions the 
experimentally controversial situation may reflect the quite 
unusual mode of critical behavior of the $N$-vector chiral model 
under the physical values of $N$. In particular, it was shown that 
this critical behavior is governed by a stable fixed point which 
is a focus, attracting RG trajectories in a spiral-like manner. 

For large enough $N$ all the field-theoretical methods are expected 
to be qualitatively and quantitatively correct. 
In fact, both $\epsilon$ and $1/N$ expansions lead to the same 
structure of flow diagram and the fixed points location and to 
identical critical exponents \cite{prv-01n}~(see also the comparison 
between Monte Carlo data and the results of three-loop FT study
made in Ref. \cite{lsd-00}). However, no analysis of the 
three-dimensional six-loop series~\cite{prv-01p} for general $N$ 
is still available.
For this reason, we calculate here all the fixed points of Hamiltonian
(\ref{LGWH}) for various $N$ and analyze the critical behavior 
characterized by the stable fixed point, computing also the chiral 
exponents on the base of known RG series \cite{prv-02}). We show 
that the focus-driven critical behavior found previously for 
physical models in Ref. \cite{cps-02} ends, under increasing $N$, 
at some $N_{c2}$. Then a small zone of first order phase transition 
$N_{c2}<N<N_{c1}$ exists. Finally, for $N>N_{c1}$ the continuous 
transition recovers, being governed by a stable node that is 
generically related to the stable fixed point given by 
$\epsilon$ and $1/N$ expansions.

The paper is organized as follows. In Sec. \ref{sec2} we describe 
the method of our analysis. Sec. \ref{sec3} contains all the 
results obtained along with their discussion. In Sec. \ref{concl} 
we present a summary and concluding remarks.

\section{Analysis method}
\label{sec2}

We analyze the fixed dimensional ($D=3$) perturbative six-loop series 
computed for general $N$ in Refs.~\cite{prv-01p,prv-02} by the same 
methods applied for the XY and Heisenberg frustrated models in 
Refs. \cite{prv-01p,cps-02}.
Within the fixed dimension approach~\cite{parisi} the perturbative 
series are obtained expanding the quantities of interest in powers 
of the quartic couplings and renormalizing the theory by introducing 
a set of zero-momentum conditions for two-point and four-point 
correlation functions \cite{prv-01p,prv-02,cppv-02}. 
All perturbative series are expressed in 
terms of the renormalized couplings $u$ and $v$, normalized
so that at the tree level they are proportional to the bare ones.
The fixed points of the theory are given by the common zeros of the 
$\beta$ functions defined as
\begin{equation}
\beta_u(u,v) = m \left. {\partial u\over \partial m}\right|_{u_0,v_0} 
,
\qquad
\beta_v(u,v) = m \left. {\partial v\over \partial m}\right|_{u_0,v_0} 
.
\label{bet}
\end{equation}
In the case of a continuous transition the coupling $u,v$ are driven 
toward an infrared-stable zero $u^*,v^*$ of the $\beta$ functions. 
The absence of a stable fixed point is usually considered as an 
indication of a first-order phase transition. The series for the 
critical exponents (e.g. $\nu(u,v)$ and $\eta(u,v)$) are obtained 
considering the anomalous dimensions of two-point functions 
with the insertion of operator with the right symmetry 
(see e.g. Refs.~\cite{ZJ-book,prv-02}).

We adopt here the symmetric rescaling of the coupling constants 
\begin{equation}
u \equiv  {16 \pi\over 3} \; R_{2N} \; \ub,\qquad\qquad
v \equiv   {16 \pi\over 3} \;R_{2N} \; \vb ,
\label{resc}
\end{equation} 
with $R_N = 9/(8+N)$ to obtain, as in two dimensions \cite{cp-01,cops-02}, 
finite fixed point coordinates in the limit of infinite 
dimension of the order parameter ($N\rightarrow \infty$). In the 
following we denote with $\bar{\beta}_{\bar{u}}$ and $\bar{\beta}_{\bar{v}}$
the $\beta $-functions corresponding to these rescaled couplings.

The obtained RG series are asymptotic and some resummation procedure 
is needed in order to extract accurate numerical values for the 
physical quantities. Exploiting the property of Borel summability of 
$\phi^4$ theories in two and three dimensions, we resum the divergent 
asymptotic series by a Borel transformation combined with a method 
for the analytic extension of the Borel transform. This last 
procedure can be obtained by a Pad\'e extension or by a conformal 
mapping \cite{LZ-77} which maps the domain of analyticity of the 
Borel transform (cut at the instanton singularity) onto a circle 
(see Refs.~\cite{LZ-77,ZJ-book} for details).

The conformal mapping method takes advantage of the knowledge of the 
large order behavior of the perturbative series 
$F(\ub,z)=\sum_k f_{k}(z) \ub^k $~\cite{prv-01p}
\begin{equation}
f_k(z) \sim k! \,(-a(z))^{k}\, k^b \,\left[ 1 + O(k^{-1})\right] 
\qquad
{\rm with}\qquad a(z) = - 1/\ub_b(z),
\label{lobh}
\end{equation}
where $\ub_b(z)$ is the singularity of the Borel transform closest 
to the origin at fixed $z= \bar{v}/\bar{u}$, given by \cite{prv-01p}
\begin{equation}
{1\over\ub_b(z)}=-a R_{2N} \quad \mbox{ for } \quad 0<z<4, 
\end{equation}
where $a=0.147\,774\,22\dots$. Moreover, it is known \cite{prv-01p} 
that for $z>2$ the Borel transform has a singularity on the 
positive real axis, which, however, is not the closest one for $z<4$. 
Thus, for $z>2$ the series is not Borel summable.

For each perturbative series $R(\bar{u},\bar{v})$,
we obtain the following approximants
\begin{equation}
E({R})_p(\alpha,b;\ub,\vb)= \sum_{k=0}^p 
  B_k(\alpha,b;\vb/ \ub) \times   \int_0^\infty dt\,t^b e^{-t} 
  {y(\ub t;\vb/\ub)^k\over [1 - y(\ub t;\vb/ \ub)]^\alpha},
\label{approx}
\end{equation}
where 
\begin{equation}
y(x;z) = {\sqrt{1 - x/\overline{u}_b(z)} - 1\over
 \sqrt{1 - x/\overline{u}_b(z)} + 1},
\end{equation}
and the coefficients $B_k$ are determined by the condition that the 
expansion of $E({R})_p(\alpha,b;\ub,\vb)$ in powers of $\ub$ and 
$\vb$ 
gives $R(\ub,\vb)$ to order $p$.

Within the second resummation procedure, the Borel transform is 
analytically extended by means of a generalized Pad\'e-Borel-Leroy 
technique, using the resolvent series trick. Explicitly, once 
introduced the resolvent series of the perturbative 
one $R(\bar{u},\bar{v})$
\begin{equation}
\tilde{P}(R)(\ub,\vb,b,\lambda)=\sum_{n} \lambda^n\sum_{k=0}^n 
{\tilde{P}_{k,n} \over (n+b)!} \ub^{n-k}\vb^{k}\, , 
\end{equation}
which is a series in powers of $\lambda$ with coefficients being 
uniform polynomials in $\ub,\vb$, the analytical continuation of the 
Borel transform is the Pad\`e approximant $[N/M]$ in $\lambda$ at 
$\lambda=1$. Obviously, the approximant for each perturbative series 
depends on the chosen Pad\'e approximant and on the parameter $b$.

In order to evaluate the fixed point coordinates by the conformal 
mapping method, we resum the perturbative series for each 
$\bar{\beta}$ function within the whole plane of coupling constants 
$(\bar{u},\bar{v})$. For each $\bar{\beta}$ function we choose 
18 approximants which were found to be stable, for small $\bar{u}$ 
and $\bar{v}$, under variation of the order of RG approximation 
(number of loops), i. e. $\alpha = 0, 2, 4$ and $b= 3,6,9,12,15,18$.
We divide the domain $0 \leq \bar{u} \leq 4$, $0 \leq \bar{v} \leq 6$ 
in $40^2$ rectangles and we mark all the sites in which at least two 
approximants for $\bar{\beta}_{\ub}$ and $\bar{\beta}_{\vb}$ vanish.
Analogously, with the Pad\'e approximants we take all of them with 
integer $b$ in the range $b\in [0,3]$ and without dangerous poles; 
in this case the statistics is much less significant due to a 
large number of defective approximants.
 
Located the fixed points, the next step is to find the stable one 
which drives the system to criticality.
The stability properties of the fixed points are determined by the 
eigenvalues $\omega_1$, $\omega_2$ of the $\Omega$ matrix
\begin{equation}
\label{omega}
\Omega = \left(\matrix{\displaystyle \frac{\partial 
\beta_u(u,v)}{\partial u} 
&\displaystyle \frac{\partial \beta_u(u,v)}{\partial v}
 \cr \displaystyle \frac{\partial \beta_v(u,v)}{\partial u} 
& \displaystyle  \frac{\partial \beta_v(u,v)}{\partial v}}\right)\; .
\end{equation}
A stable fixed point must have two eigenvalues with positive real 
parts, while a fixed point possessing eigenvalues with non-vanishing 
imaginary parts is called focus.

We calculate $\omega_1$ and $\omega_2$ at a given fixed point 
considering the numerical derivatives of each pair of approximants 
of the two $\bar{\beta}$-functions at their common zero (there are 
324 possible combinations with the conformal mapping method and 81, 
minus defective ones, with the Pad\'e approximant technique).
To obtain reliable numerical estimates for $\omega_1$ and $\omega_2$, 
we limit ourselves by the approximants which yield the fixed point 
coordinates compatible with their final, properly weighted values.

The critical exponents are finally evaluated at the stable fixed 
point by means of employing of the approximants which minimize the 
difference between the estimates given by different perturbative 
orders~(see e.g. \cite{cpv-00,pv-00} for details).

\section{Results}
\label{sec3}

\begin{figure}[t]
\centerline{\epsfig{height=16.6truecm,width=12truecm,file=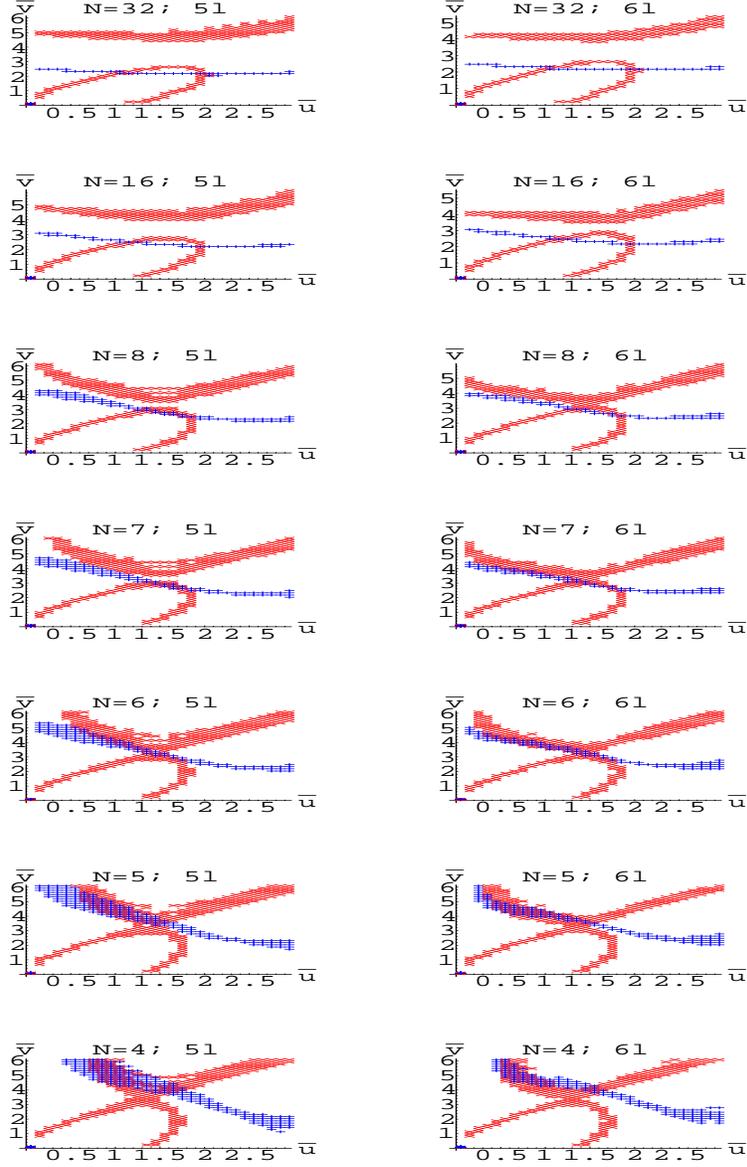}}
\caption{Zeros of $\bar{\beta}$-functions for  
$N=32,\,16,\,8,\,7,\,6,\,5$, and $4$ in the $(\bar{u},\bar{v})$ 
plane.
Pluses ($+$) and crosses ($\times$) correspond to zeroes of 
$\bar{\beta}_{\bar{v}}(\bar{u},\bar{v})$ and 
$\bar{\beta}_{\bar{u}}(\bar{u},\bar{v})$ respectively.}\label{alln}
\end{figure}

In this section we analyze the six-loop perturbative series of the RG 
functions for large number of components of the order parameter using  
the methods explained above. In Fig.\ref{alln} we 
report our results for the zeros of the two $\bar{\beta}$ functions 
for $N=32,\,16,\,8,\,7,\,6,\,5$, and $4$, obtained with the conformal 
mapping technique. It is seen from this figure that two branches 
of zeros of $\bar{\beta}_{\bar{u}}$ exist for all the considered 
values of $N$. These branches become closer to each other with 
decreasing $N$. The zeros of $\bar{\beta}_{\bar{v}}$ are 
represented by a curve which clearly intersects the upper branch 
of $\bar{\beta}_{\bar{u}}$ for $N \leq 5$ and the lower branch for 
$N\geq 7$. In the complement of these regions, i.e. for $N=6$ and 
close vicinity of this value, the presence or the absence of fixed 
points is less evident from Fig.\ref{alln}. However, looking at 
the single pair of approximants for two $\bar{\beta}$ functions, 
one can find that for all $5<N<7$ the majority of the approximants 
for $\bar{\beta}_{\bar{u}}$ gives rise to two branches of zeros 
separated in the plane $(\ub,\vb)$.

For each pair of approximants we find that two marginal values of $N$, 
$N_{c1}$ and $N_{c2}$, exist. For $N>N_{c1}$ the chiral and the 
antichiral fixed points are always present. They are located on the 
lower branch of $\bar{\beta}_{\bar{u}}$ and coalesce for $N=N_{c1}$.
For $N_{c2}<N<N_{c1}$ only the unstable -- Gaussian and 
O$(2N)$-symmetric (Heisenberg) -- fixed points exist, while for 
$N\leq N_{c2}$ the chiral and the antichiral fixed points appear 
again, being located, however, on the upper branch of 
$\bar{\beta}_{\bar{u}}$. With the precision of our calculation we 
can state that $N_{c1}= 6.4(4)$ and $N_{c2}= 5.7(3)$.

Let us now discuss the critical behavior in these three domains of $N$.
For $N \geq N_{c1}$ there are four fixed points: chiral, antichiral, 
Heisenberg and Gaussian. The location (reported in Table~\ref{tabfp}) 
and the stability analysis~(see Table~\ref{tabellone2}) of these 
fixed points reveal that the chiral one is a stable node and governs 
the critical regime, while the Heisenberg and the Antichiral are 
unstable with one negative eigenvalue of the $\Omega$ matrix~(cf. 
Table~\ref{tabellone2}). The critical exponents that characterize 
the second order phase transition are reported in 
Tab.~\ref{tabellone2} with the data of the large $N$ analysis~
\cite{prv-01p,g-02n,g-02p}. The results indicate that there is a 
reasonable agreement between the two perturbative methods. It implies 
that this domain of $N$ is generically related to those accessible 
in the framework of the $1/N$ and $\epsilon$ expansions. So, we can 
conclude that $N_{c1}$ is nothing but a marginal value of $N$ 
introduced earlier within various approximations ($N_{c3}$ of 
Refs.~\cite{asv-95,as-94}, $n_I (d)$ of Ref.~\cite{Kawamura-98}, 
$n^{+}(m,d)$ of Ref.~\cite{prv-01n}, $N_c$ of Ref.~\cite{prv-01p}, 
and $N_{+}$ of Ref.~\cite{rev-01}).

\begin{table}[t]
\squeezetable
\caption{
Location of the chiral and antichiral fixed point as function of 
$N$.}
\begin{tabular}{l|c|c|c||c|c}
$N$ & Ch.(5loop) & Ch.(6loop) & Ch.(Pad\'e) & ACh.(5loop) & ACh.(6loop)  
\\
    &  $(\bar{u}^*_{ch},\bar{v}^*_{ch})$ & 
$(\bar{u}^*_{ch},\bar{v}^*_{ch})$
    & $(\bar{u}^*_{ch},\bar{v}^*_{ch})$
    & $(\bar{u}^*_{ach},\bar{v}^*_{ach})$ & 
$(\bar{u}^*_{ach},\bar{v}^*_{ach})$ \\
\tableline \hline
 2    & [1.87(13),5.33(15)]   & [1.9(1),5.47(20)]\cite{prv-01p}   & 
[1.9(2),5.25(35)] &  & \\
 3    & [1.8(1),4.65(30)]     & [1.8(1),4.67(23)]\cite{prv-01p}   & 
[1.74(13),4.35(11)]&& \\
 4    & [1.59(15),4.0(4)]   & [1.54(13),3.96(21)] & 
[1.62(10),3.84(6)] && \\ 
 5    & [1.48(12),3.58(28)]   & [1.44(15),3.60(22)] & 
[1.50(5),3.40(10)] && \\ 
 6 & --  & -- & -- & -- & --\\
 7    & [1.72(5),2.69(7)] & [1.75(4),2.67(5)] & 
[1.756(10),2.648(12)] &
[1.43(11),3.01(14)] & [1.39(11),3.05(12)]\\
 8    & [1.81(3),2.51(5)] & [1.82(2),2.50(2)] & 
[1.826(8),2.491(13)] &
[1.39(11),2.89(11)] & [1.33(8),2.93(8)]\\
16 & [1.963(12),2.201(10)] &[1.968(6),2.204(4)]     & 
[1.967(3),2.205(5)] &
[1.178(23),2.470(11)] & [1.156(17),2.479(9)] \\
32 & [2.000(4),2.1027(29)] & [2.000(1),2.103(3)]    & 
[2.000(4),2.101(2)] &
[1.086(6),2.2381(15)] &[1.075(5),2.2406(12)]\\ 
\end{tabular}
\label{tabfp}
\end{table}

For $N_{c2} < N < N_{c1}$ no stable fixed points exist in the region 
$\bar{u}>0,\bar{v}>0$ and the transition is expected to be first order~
(e.g. for $N=6$ we found, at the $O(2N)$-symmetric fixed point, 
$\omega_{O(12)}=-0.53(2)$).


For $N<N_{c2}$, as seen from Fig. \ref{alln}, the chiral fixed point
lays on the upper branch of zeros of the $\bar{\beta}_{\bar{u}}$, so 
in principle there is no reason to obtain the results related to those 
of the large $N$ approach. In this domain, the transition is continuous 
and the eigenvalues of the $\Omega$ matrix at the chiral fixed point 
possess nonvanishing imaginary parts in the majority of the cases. 
In Table~\ref{tabellone} we report the estimates for $\omega_1$ and 
$\omega_2$ and the percentage of complex eigenvalues. 
Thus the critical behavior is driven by a stable focus.
In Table~\ref{tabellone} we report the critical exponents obtained 
in this article along with the already known values for $N=2,3$~
\cite{prv-01p,prv-02} in order to make visible all the trends of 
physical quantities with varying $N$.
We also find that for almost all the working approximants, 
the antichiral fixed point is located in the region where the 
singularity of the Borel transform closest to the origin is on the 
real positive axis, leaving its existence doubtful. The fact
that its position strongly oscillates with varying the
approximants indicates that in this domain of $(\bar{u}, \bar{v})$ 
plane the analysis is not robust. On the other hand, the
presence of the stable focus chiral point does not imply the 
existence of the antichiral one. The unstable fixed point is not 
topologically needed on the separatrix dividing the regions of the 
first-order and second-order phase transitions.

\begin{table}[b]
\squeezetable
\caption{
Critical behavior for $N>N_{c1}$. Values obtained with other methods 
are reported for comparison. 
In the last line, only the positive eigenvalues at the antichiral fixed 
point are reported, since the negative ones are not obtainable with 
the method of Ref.~\protect \cite{g-02n}.
}
\begin{tabular}{l|c|c|c|c}
Eigenvalues $\omega_1$, $\omega_2$ & $N=7$ & $N=8$ & $N=16$ & $N=32$ \\
\hline
C.M.(5loop)& [0.84(2),0.21(8)] & [0.84(2),0.33(7)] & 
[0.878(6),0.702(13)]& [0.930(3),0.863(5)]    \\
C.M.(6loop)& [0.83(2),0.23(5)]  & [0.83(2),0.36(4)] & 
[0.876(4),0.714(9)] & [0.933(2),0.868(2)]    \\
Pad\'e     & [0.84(1),0.24(2)] &   [0.82(3),0.35(3)] & 
[0.87(3),0.70(3)   & [0.92(2),0.83(4)] \\
O$(1/N)$ \cite{g-02n}&[0.768, 
0.537]&[0.797,0.594]&[0.899,0.797]&[0.949,0.899] \\
\tableline \hline
$\nu$ & $N=7$&$N=8$ &$N=16$ & $N=32$   \\
\hline
C.M.    &0.68(2)    & 0.71(1)  & 0.863(4) & 0.936(1)   \\
Pad\'e  & 0.65(4)   & 0.72(1) &--  & --  \\
O$(1/N^2)$ \cite{prv-01n} & 0.697   & 0.743 & 0.885 & 0.946  \\
\tableline \hline
$\gamma$& $N=7$&$N=8$ &$N=16$ & $N=32$  \\
\hline
 C.M.  & 1.31(5) & 1.40(2) & 1.68(1) & 1.830(6) \\
$O(1/N^2)$ \cite{prv-01n} & 1.36   & 1.45 & 1.75 & 1.88  \\
\tableline \hline
$\eta$& $N=7$&$N=8$ &$N=16$ & $N=32$   \\
\hline
 C.M.  & 0.042(2) & 0.039(1) & 0.0246(2) & 0.01357(1) \\
Pad\'e     &  0.040(3)  & 0.038(2)  & 0.0245(5) & 0.01357(3)    \\
O$(1/N^2)$  \cite{prv-01n}& 0.053   & 0.04724 & 0.0245 & 0.01245   \\
\tableline \hline
$\gamma_c$& $N=7$&$N=8$ &$N=16$ & $N=32$   \\
C.M. &0.74(2)  &0.77(1) &0.867(4)&0.918(2) \\
Pad\'e &0.75(2)  & 0.76(2)&--& -- \\
O$(1/N^2)$ \cite{g-02p}& 0.765 & 0.790 & 0.890 & 0.945\\
\tableline \hline
$\beta_c$& $N=7$&$N=8$ &$N=16$ & $N=32$   \\
\hline
C.M. & 0.64(2) & 0.69(1)&0.840(5)& 0.927(2)\\
O$(1/N^2)$ \cite{g-02p}& 0.755 &0.781 & 0.885& 0.944 \\
\tableline \hline
$\omega_{\rm ach}$ & $N=7$&$N=8$ &$N=16$ & $N=32$  \\
\hline
C.M.(5loop)& [0.88(5),-0.23(8)] & [0.91(4),-0.37(7)]  & 
[0.973(6),-0.760(16)]& [0.9882(12),-0.899(6)]     \\
C.M.(6loop)& [0.87(4),-0.23(4)]  & [0.90(3),-0.38(2)] & 
[0.969(6),-0.762(9)] & [0.9864(11),-0.900(4)]   \\
$O(1/N^2)$ \cite{g-02n}& [0.923,--] &[0.932,--] & [0.966,--]& 
[0.983,--] \\
\end{tabular}
\label{tabellone2}
\end{table}

In the Tables~\ref{tabfp}, \ref{tabellone2}, and \ref{tabellone} we 
report, along with the results given by the conformal mapping 
technique commented above, all the estimates obtained with the 
Pad\'e-Borel resummation method. Using this method, we faced with 
a lot of defective Pad\'e approximants which in some cases prevent 
us from finding estimates for physical quantities (e.g. the values 
of $\nu$ for $N=16$ and 32), while in other ones turn out to lead to 
underestimation of the apparent errors. Anyway, the data obtained 
with the two resummation techniques are found to be substantially 
equivalent.

\begin{figure}[t]
\centerline{\epsfig{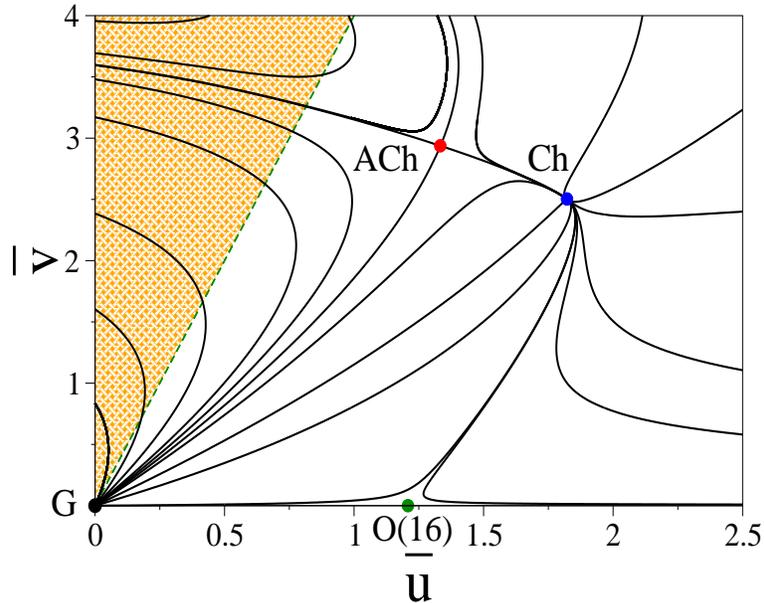}}
\caption{RG flow in the renormalized coupling constant plane for 
$N=8$.}
\label{rgn8}
\end{figure}
\begin{figure}[t]
\centerline{\epsfig{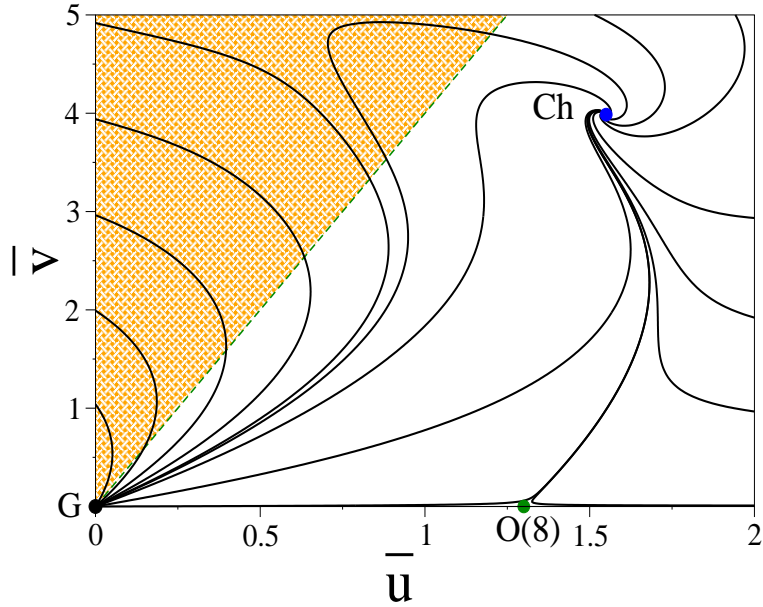}}
\caption{RG flow in the renormalized coupling constant plane for 
$N=4$.}
\label{rgn4}
\end{figure}

In Figs.~\ref{rgn8},\ref{rgn4} we display two example of the RG flows 
given by typical working approximants for $N=8$ and $N=4$ respectively. 
These flows clearly demonstrate that for $N=4$ there is spiral-like 
approach to the stable fixed point. This is no longer true for $N=8$.
Obviously, all the RG flows are quantitatively correct only within 
the regions where the singularity of the Borel transform closest 
to the origin is on the real negative axis~\cite{prv-01p}; in 
Figs.~\ref{rgn8},\ref{rgn4} they correspond to unshaded areas. 
Nevertheless, we report also the flows in other parts of the 
coupling constant plane to present a complete qualitative picture.
The region that is outside the domain of attraction of the stable 
fixed point corresponds to the domain of the first-order phase 
transitions. Note, that saying ``the domain of the first-order phase
transitions'' we mean, as usually, the region where the quartic form 
in free-energy expansion can acquire negative values. In fact, 
because of the presence of the higher-order terms in this expansion 
that make the system globally stable at any temperature, the true 
domains of the first-order transitions may be substantially more 
narrow than those predicted by the mean-field approximation.

\begin{table}[t]
\squeezetable
\caption{Critical behavior in the region $N<N_{c2}$.
Real eig. (Imaginary eig.) stays for real eigenvalues (imaginary 
eigenvalues). In the lines marked by \% Real we report the 
percentages of the purely real eigenvalues. Close to the values 
obtained in other works one can find the original references.}
\begin{tabular}{l|c|c|c|c}
Real eig.       & $N=2$  &$N=3$     &$N=4$   &   $N=5$    \\
\hline
C.M.(5loop)&     &&   & [0.63(18),0.24(16)]     \\
\% Real    & 2.5\%   &         &           & 35\% \\
C.M.(6loop)&   &  &    & [0.66(16),0.30(17)]  \\
\% Real    & 23.5\%  &  11\%  &      2\%       &   24\%      \\
\tableline\hline
Imaginary eig. & $N=2$   &$N=3$ &$N=4$   &  $N=5$ \\
\hline
C.M.(5loop)&  1.40(35)$\pm i$1.00(45) \cite{cps-02}  &0.85(20)$\pm 
i$0.95(15) \cite{cps-02}& 0.58(11)
$\pm i$0.52(25) & 0.43(10)$\pm i$0.25(22) \\
C.M.(6loop)& 1.70(40)$\pm i$0.90(40) \cite{cps-02}& 1.00(20)$\pm 
i$0.80(25) \cite{cps-02}& 0.67(11)
$\pm i$0.53(23) & 0.49(6)$\pm i$0.22(19) \\
\tableline \hline
$\nu$ & $N=2$&$N=3$ &$N=4$ & $N=5$ \\
\hline
C.M.  & 0.57(3) \cite{prv-01p}  & 0.55(3) \cite{prv-01p}&0.51(3)  
&0.51(3)  \\
Pad\'e     &  0.64(25)   & 0.59(13)  & 0.54(12) & 0.52(11)   \\
\tableline \hline
$\gamma$& $N=2$&$N=3$ &$N=4$ & $N=5$ \\
\hline
 C.M.  & 1.11(5) \cite{prv-01p}& 1.05(5) \cite{prv-01p}& 1.00(5) & 
1.02(5)\\
\tableline \hline
$\eta$& $N=2$&$N=3$ &$N=4$ & $N=5$ \\
\hline
 C.M.  & 0.095(7) \cite{prv-01p}& 0.088(9) \cite{prv-01p} & 0.066(6) 
& 0.056(6) \\
Pad\'e     &  0.095(9)  & 0.086(11)  & 0.069(13) & 0.060(12)   \\
\tableline \hline
$\phi_c$& $N=2$&$N=3$ &$N=4$ & $N=5$ \\
\hline
 C.M.  & 1.43(4) \cite{prv-02}& 1.27(4) \cite{prv-02} &1.17(4) 
&1.12(5)\\
Pad\'e &1.4(1) &1.23(10)&1.15(10)&1.08(12)\\
\tableline \hline
$\beta_c$& $N=2$&$N=3$ &$N=4$ & $N=5$ \\
\hline
 C.M.  & 0.28(10) \cite{prv-02}& 0.38(10) \cite{prv-02} & 
0.52(3)&0.51(3)\\
\end{tabular}
\label{tabellone}
\end{table}

\section{Conclusion}
\label{concl}

In this work, we have studied the critical thermodynamics of the 
O$(N)\times $O$(2)$ spin model for number of components of the 
order parameter $N$ greater than $3$ on the base of the six-loop RG 
series~\cite{prv-01p,prv-02}, giving a complete description of the 
evolution of critical behavior under varying $N$. We have found 
that for $N>N_{c1}\sim 6.4(4)$ the transition is continuous 
and governed by a stable chiral fixed point, that is generically 
related to the stable fixed point found earlier within the 
$\epsilon$ expansion and $1/N$ expansion analyses.
For $N_{c2}<N<N_{c1}$~with $N_{c2}=5.7(3)$ we have not detected 
any stable fixed point and the chiral transition is expected to be 
first order. The focus-like critical behavior, found previously for 
physical models ($N=2,3$) in Ref. \cite{cps-02}, has been confirmed to 
exist for any $N<N_{c2}$. In this region of $N$, the character of 
crossover phenomena may be rather complicated leading to effective 
exponents which vary in a non-monotonic way. In analogy with dilute 
Ising systems, where the concentration-dependent critical exponents 
seem to violate the universality because of crossover effects near 
criticality~\cite{jos-95}, it looks plausible that the results of some 
simulations~\cite{sim} and experiments~\cite{esp}, which yield negative 
values for the exponent $\eta$, reflect the peculiarities of the 
preasymptotic regimes.

Being the region $N>N_{c1}$ connected with that accessible by the 
large $N$ analysis, we expect that $N_{c1}$ coincides with $N_c$ 
usually introduced in $\epsilon$ and $1/N$ expansions. Several 
estimates for $N_c$ exist. To quote the more recent ones, we have 
$N_{c}=5.3(2)$ \cite{prv-01n} and $N_c\sim3.39$ \cite{asv-95} from 
3-loop $\epsilon$ expansion, $N_{c}\sim 5$ 
 from the exact renormalization 
group approach \cite{tdm-00}, 
and $N_{c}=3.24$ \cite{g-02n} and $N_{c}=5.3$  \cite{prv-01n}
from $1/N$ expansion. 
These scattered results clearly indicate that the extrapolation of 
$1/N$ and $\epsilon$ expansion predictions up to the physical values 
of $N$ and $\epsilon$ is a quite delicate matter. 

To conclude, we believe that working with the highest-order, six-loop 
RG series and processing them by means of the advanced resummation 
techniques, we obtain a clear and definite understanding of the 
critical behavior of $N$-vector chiral model for all the integer 
values of $N$, including the physically relevant ones.

\section*{ACKNOWLEDGMENTS}
We would like to thank Ettore Vicari and Andrea Pelissetto for many 
useful discussions. The financial support of the Russian Foundation 
for Basic Research under Grant No. 01-02-17048 (A.I.S.) and the 
Ministry of Education of Russian Federation under Grant 
No. E02-3.2-266 (A.I.S.) is gratefully acknowledged. A.I.S. has 
benefited from the warm hospitality of Scuola Normale Superiore and 
Dipartimento di Fisica dell'Universit\`a di Pisa, where this research 
was done.


\end{document}